\documentclass[12pt,preprint]{aastex}

\def\kms{\ifmmode{\rm km\thinspace s^{-1}}\else km\thinspace s$^{-1}$\fi}
\def\ms{\ifmmode{\rm m\thinspace s^{-1}}\else m\thinspace s$^{-1}$\fi}

\shortauthors{Konacki et al.}
\shorttitle{OGLE-TR-113}

\begin{document}

\title{A transiting extrasolar giant planet around the star OGLE-TR-113}

\author{Maciej Konacki\altaffilmark{1}, Guillermo Torres\altaffilmark{2},
Dimitar D. Sasselov\altaffilmark{2}, Grzegorz Pietrzy\'nski\altaffilmark{3,4},
Andrzej Udalski\altaffilmark{4}, Saurabh Jha\altaffilmark{5},
Maria Teresa Ruiz\altaffilmark{6}, Wolfgang Gieren\altaffilmark{3},
and Dante Minniti\altaffilmark{7}}

\altaffiltext{1}{California Institute of Technology, Div.\ of
Geological \& Planetary Sciences 150-21, Pasadena, CA 91125, USA;
e-mail: maciej@gps.caltech.edu}

\altaffiltext{2}{Harvard-Smithsonian Center for Astrophysics, 60
Garden St., Cambridge, MA 02138}

\altaffiltext{3}{Universidad de Concepci\'on, Departamento de F\'\i
sica, Astronomy Group, Casilla 160-C, Concepci\'on, Chile}

\altaffiltext{4}{Warsaw University Observatory, AL.\ Ujazdowskie 4,
00-478, Warsaw Poland}

\altaffiltext{5}{Department of Astronomy, University of California,
Berkeley, CA 94720, USA}

\altaffiltext{6}{Departamento de Astronom\'\i a, Universidad de Chile,
Casilla 36-D, Santiago, Chile}

\altaffiltext{7}{Pontificia Universidad Cat\'olica de Chile,
Departamento de Astronom\'\i a y Astrof\'\i sica, Casilla 306,
Santiago 22, Chile}

\begin{abstract}
We report the independent discovery of a new extrasolar transiting planet around
OGLE-TR-113, a candidate star from the Optical Gravitational Lensing
Experiment. Small radial-velocity variations have been detected based
on observations conducted with the MIKE spectrograph on the
Magellan~I (Baade) telescope at the Las Campanas Observatory (Chile)
during 2003. We have also carried out a light-curve analysis
incorporating new photometry and realistic physical parameters for the
star.  OGLE-TR-113b has an orbital period of only 1.43 days, a mass of
$1.08\pm0.28$~M$_{\rm Jup}$, and a radius of $1.09\pm0.10$~R$_{\rm
Jup}$. Similar parameters have been obtained very recently in an
independent study by Bouchy et al., from observations taken a year
later.  The orbital period of OGLE-TR-113b, and also that of the
previously announced planet OGLE-TR-56b ($P_{\rm orb}=1.21$~days)
---the first two found photometrically--- are much shorter than the
apparent cutoff of close-in giant planets at 3-4-day periods found
from high-precision radial velocities surveys. Along with a third case
reported by Bouchy et al.\ (OGLE-TR-132b, $P_{\rm orb} = 1.69$~days),
these objects appear to form a new class of ``very hot Jupiters'' that
pose very interesting questions for theoretical study. 

 \end{abstract}


\section{Introduction}

In recent years the field of extrasolar planet research has seen
significant developments in the ability to discover and measure these
objects using a variety of techniques. High-precision Doppler searches
\citep[e.g.,][]{Fischer:03, Naef:04} have yielded the vast majority of
the discoveries, and measurements in at least one case have even been
made astrometrically \citep{Benedict:02}. Transit searches had their
first success with HD~209458 \citep{Henry:00, Charbonneau:00}, a
bright star ($V = 7.65$) that was known previously to harbor a planet
in a 3.5-day period orbit from its radial-velocity signature.
Numerous photometric programs are monitoring large samples of stars
looking for small dips in the brightness of the central object at the
$\sim$1\% level \citep[see][]{Horne:03}, which might indicate a
planet-size object crossing in front of the star.  These studies are
very important for the additional information they bring to bear on
the nature of the companion, namely, the inclination angle of the
orbit ($\sim$90\arcdeg) and the absolute radius of the planet.  The
inclination angle complements the spectroscopic information and allows
a direct determination of the mass.

Dozens of transiting planet candidates among faint stars have already
been reported by several teams including OGLE \citep{Udalski:02a},
EXPLORE \citep{Mallen-Ornelas:03}, MACHO \citep{Drake:04}, and others.
Multiple efforts are underway to follow-up on these candidates, a
necessary step given the high incidence of false positive detections,
particularly among fainter stars in crowded fields.  The first case to
be confirmed was that of OGLE-TR-56, a star with $V = 16.6$ located in
the direction of the Galactic center \citep{Udalski:02b, Konacki:03a,
Torres:04}. The very short orbital period of this planet (only
1.21~days) makes it extremely interesting, and has provided theorists
the opportunity to explore the effects of strong irradiation from the
central star as well as evaporation \citep[e.g.,][]{Burrows:03,
Baraffe:03, Baraffe:04}.

In this paper we report the detection of a Doppler signature induced
by a giant planet orbiting OGLE-TR-113, another faint transit
candidate ($I = 14.42$) in the constellation of Carina reported
recently by the OGLE project \citep{Udalski:02c}. This star shows
periodic dips in brightness of about 3\%, and has a photometric period
of 1.43~days.  OGLE-TR-113 was originally identified as a very
promising candidate from our low-res\-o\-lu\-tion spectroscopic
observations conducted in 2002 \citep[see][]{Konacki:03b}. This
reconnaissance showed it to be a star of late spectral type with no
obvious velocity variations at the level of a few \kms, which would
have otherwise disqualified it for implying a stellar
companion. Subsequently it was placed on our program for
high-resolution follow-up, and the observatios were carried out in
early 2003. As this paper was being prepared we learned of a very
recent independent detection of radial velocity variations in
OGLE-TR-113 by \cite{Bouchy:04}, based on observations taken in 2004. 
That study found yet another case of a very short-period
transiting planet (OGLE-TR-132, $P_{\rm orb} = 1.69$~d), which brings
the number of such objects to three. It appears, therefore, that they
form a new class of ``very hot Jupiters'' not previously seen in
high-precision radial-velocity surveys.

\section{Observations and reductions}
\label{sec:observations}

Our high-resolution spectroscopic observations were carried out with
the MIKE spectrograph \citep{Bernstein:03} on the Magellan~I (Baade)
telescope at the Las Campanas Observatory in Chile. Seven spectra of
OGLE-TR-113 were obtained from February to April of 2003.  The
resolving power of these observations is $\lambda/\Delta\lambda
\approx 54,000$, and the wavelength coverage is from 450 to
725~nm. Only 22 of the 29 echelle orders were used, since the others
had low signal or were affected by telluric lines. The average
signal-to-noise ratios achieved in our 30--40~minute exposures range
from about 15 to 20 per pixel. In addition to OGLE-TR-113 we observed
several brighter stars with known planets each night as radial
velocity standards. The wavelength reference for all observations was
determined from exposures of a hollow-cathode Thorium-Argon lamp taken
immediately before and after each stellar exposure.

The spectra were reduced using standard tasks in IRAF\footnote{IRAF is
distributed by the National Optical Astronomy Observatories, which is
operated by the Association of Universities for Research in Astronomy,
Inc., under contract with the National Science Foundation.}, as well
as rectification tools developed by \cite{Kelson:03}.  Radial
velocities were obtained by cross-correlation against a calculated
template that was tuned to match the star.  For OGLE-TR-113 the
stellar parameters we determined are $T_{\rm eff} = 4800\pm150$~K,
$\log g = 4.5^{+0.5}_{-0.8}$, [Fe/H] $= 0.0^{+0.1}_{-0.3}$,
macroturbulent velocity $\zeta_{\rm RT} = 2\pm1$~\kms, and $v \sin i =
9\pm3$~\kms. These were derived by careful comparison of calculated
LTE model spectra against features such as the H$\alpha$, H$\beta$,
and Na~D lines, in addition to numerous other metal lines, which taken
together provide strong constraints on the effective temperature and
surface gravity of the star. The parameters for the standards were
adopted from detailed analyses in the literature.  The velocity
results from the different orders were combined for each star, and the
scatter between orders was used to derive an estimate of the
uncertainty.  Instrumental shifts during the night were monitored and
corrected for by using telluric lines present in the spectrum. Typical
corrections average 100-150~\ms\ with occasionally larger values, and
they are found to improve the accuracy of the velocities
significantly.

\begin{figure}
\figurenum{1}
\hskip 0.9in \includegraphics[scale=0.85]{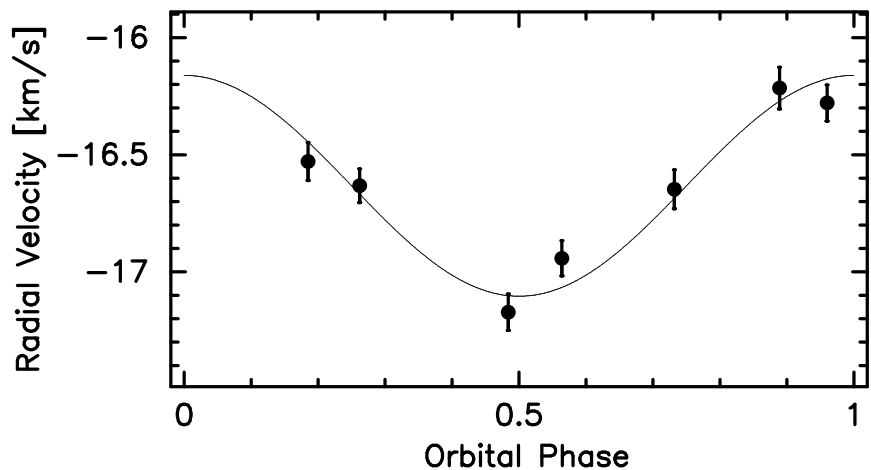}
 \caption{Radial velocity measurements for $\tau$~Boo as a function of
orbital phase, along with the orbit determined by \cite{Butler:97}.
The only parameter we have adjusted is the velocity of the center of
mass.\label{fig:tauboo}}
 \end{figure}

The results for one of our standards, $\tau$~Boo (HD~120136), are
shown in Figure~\ref{fig:tauboo}.  The measured radial velocities are
in excellent agreement with the known spectroscopic orbit for this
star from \cite{Butler:97}, which has a semi-amplitude of about
470~\ms.  The RMS residual from the published orbit is 85~\ms, and the
only parameter adjusted to match the observations is a velocity
offset.  This demonstrates that the instrumental setup allows us to
clearly detect small velocity changes at the level of a few hundred
\ms.

\section{Spectroscopic results}
\label{sec:sborbit}

\begin{figure}
\figurenum{2}
\hskip 0.9in \includegraphics[scale=0.85]{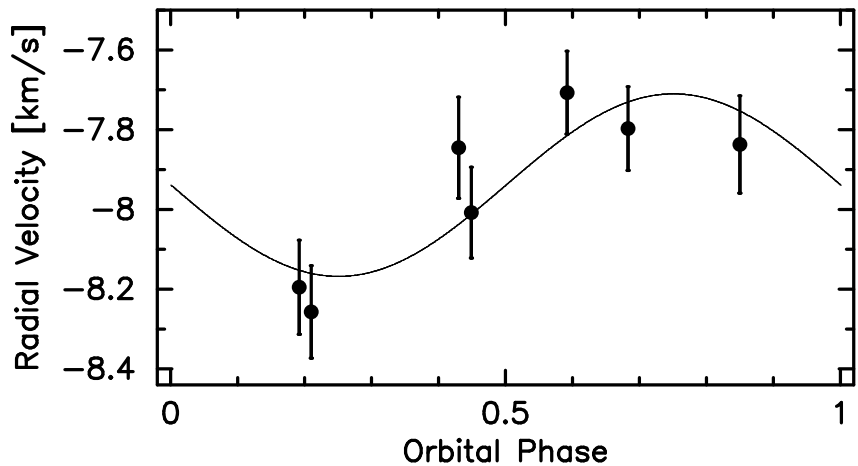}
 \caption{Radial velocity measurements and fitted velocity curve for
OGLE-TR-113, as a function of orbital phase. Only the semi-amplitude
and center-of-mass velocity have been adjusted. The transit ephemeris
is adopted from the photometry (see text)\label{fig:oglervs}.}
 \end{figure}

Our velocity measurements for OGLE-TR-113 are listed in
Table~\ref{tab:rvs} and shown in Figure~\ref{fig:oglervs}. Typical
measurement errors are $\sim$100~\ms. We fitted a Keplerian orbit to
these observations holding the well-determined period and epoch fixed
from the photometry (see \S\ref{sec:lightcurve}), and adjusting only
the velocity semi-amplitude ($K$) and the center-of-mass velocity
($\gamma$).  A circular orbit was assumed, based on the extremely
short orbital period and the likelihood that tidal forces have reduced
the eccentricity to zero.  We obtained $K = 229 \pm 58$~\ms\ and
$\gamma = -7.939 \pm 0.043$~\kms, with an RMS residual from the fit of
108~\ms\ (Table~\ref{tab:results}).  The semi-amplitude is significantly
different from zero and it is robust. The minimum mass of the companion 
derived from our best orbital fit is $M_p \sin i = (0.00121 \pm 0.00031)
(M_*+M_p)^{2/3}$~M$_{\sun}$, where $M_*$ is the mass of the primary
star. Formally our systemic velocity for OGLE-TR-113 agrees very well
with the value of $\gamma = -7.944 \pm 0.027$~\kms\ by
\cite{Bouchy:04}. For our standard $\tau$~Boo we obtain $\gamma =
-16.632 \pm 0.033$~\kms, which also compares favorably with the
determination by \cite{Nidever:02} of $-16.542$~\kms, refered to a
well-defined velocity system. However, we do not claim here that the
accuracy of our zero point is much better than $\sim$100~\ms, partly
because of the small number of observations. Nevertheless, the
similarity of the systemic velocity we derive for OGLE-TR-113 with
that of \cite{Bouchy:04} places at least some constraint on the
presence of additional massive planets in wider orbits around the
star.

\begin{figure} 
\figurenum{3} 
\vskip -2.2in \includegraphics[scale=0.8]{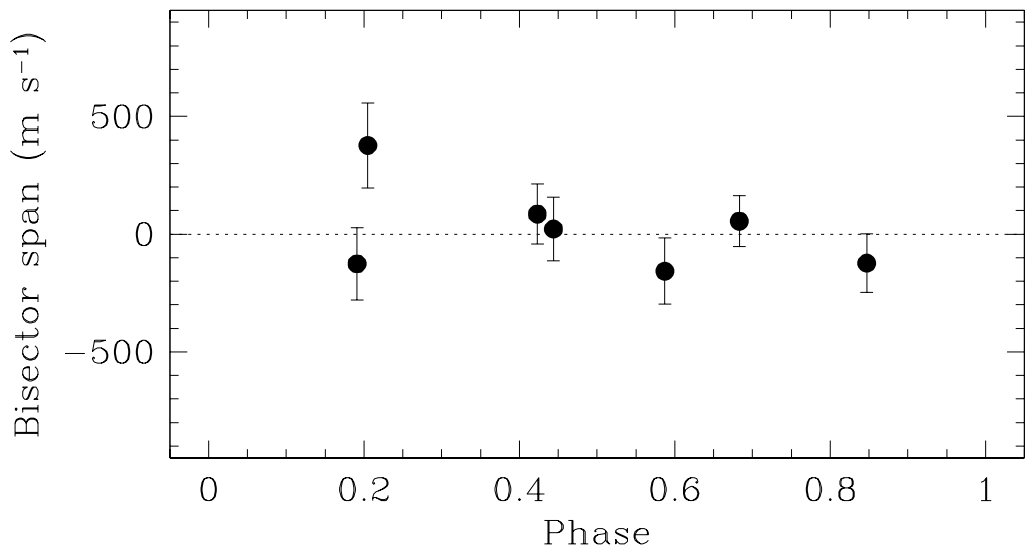}
\vskip -1.2in 
 \caption{Bisector span used to quantify the line asymmetry for each
of our spectra of OGLE-TR-113, plotted as a function of orbital phase.
The error bars are determined from the agreement between different
echelle orders. There is no obvious correlation with
phase.\label{fig:bisplot}}
 \end{figure}

Among the phenomena that can mimic the photometric and spectroscopic
signatures of transit candidates, the presence of an eclipsing binary
along the same line of sight (a ``blend") is one of the most common.
Deep eclipses in the binary can be strongly diluted by the main star,
and appear with depths of only a few percent that are very similar to
those produced by a Jupiter-size planet around a solar-type star.
Furthermore, light from one of the stars in the eclipsing binary can
contaminate the spectrum of the main star, and introduce line
asymmetries that could lead to spurious velocity variations. To
examine this possibility we quantified the asymmetries by computing
the line bisector spans \citep[see, e.g.,][]{Gray:92}. This was done
directly from the cross-correlation function of OGLE-TR-113 co-added
over all echelle orders, following \cite{Santos:02} and
\cite{Torres:04}. The results are shown as a function of phase in
Figure~\ref{fig:bisplot}. Within the errors we detect no significant
variation, indicating that line asymmetries cannot be the source of
the velocity variations ($\sim$500~\ms, peak to peak) since these two
quantities should be of the same order.
	
\section{Light curve solution}
\label{sec:lightcurve}

\begin{figure}
\figurenum{4}
\vskip 0.2in \hskip 0.2in \includegraphics[scale=0.95]{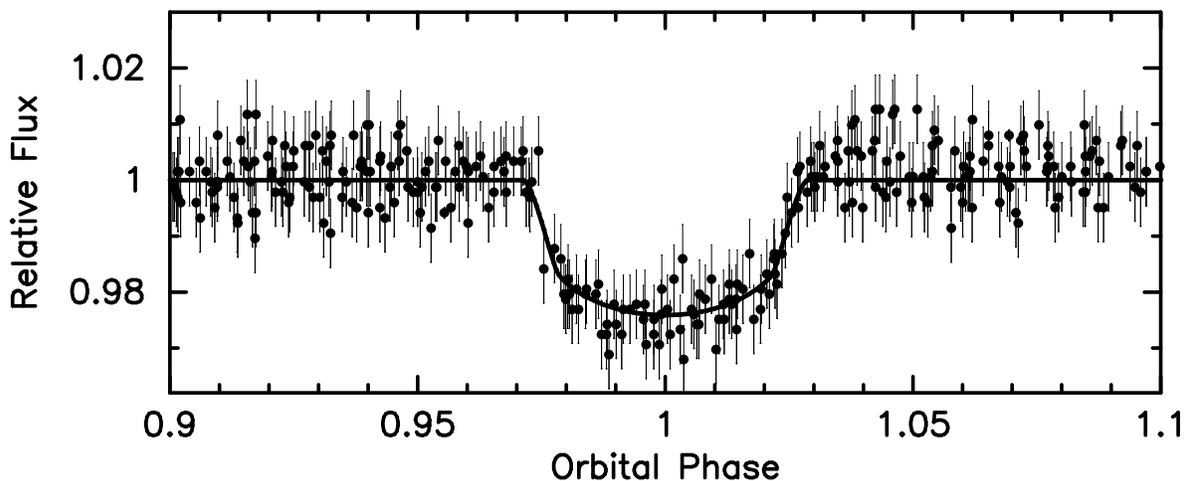}
 \caption{OGLE photometry for OGLE-TR-113 in the $I$ band, with our
best fit transit light curve. The resulting parameters are listed in
Table~\ref{tab:results}.\label{fig:lcurve}}
 \end{figure}

The discovery of a transit signature in OGLE-TR-113 was reported by
\citep{Udalski:02c}, and was based on observations collected during
the 2002 observing season in which 10 transit events were recorded.
Since then further measurements spanning two additional seasons have
been made (for a total of 1517), and 4 more transits have been
detected. We have incorporated these new measurements into our
analysis. The internal errors of 0.003~mag, possibly a bit optimistic,
were rescaled to 0.006~mag for consistency with the out-of-eclipse
variations and also to provide a reduced $\chi^2$ near unity in the
light curve solution. The best fit transit model, computed following
\cite{Mandel:02}, is shown in Figure~\ref{fig:lcurve} together with
the observations. The resulting parameters along with the period and
transit epoch, which were adjusted simultaneously, are given in
Table~\ref{tab:results}.  The improved period and transit epoch are
$P_{\rm orb} = 1.4324758 \pm 0.0000046$~days and $T {\rm (HJD)} =
2,\!452,\!325.79823 \pm 0.00082$. The limb-darkening coefficient in
the $I$ band (linear law), $u_I = 0.586\pm0.015$, was derived from
models consistent with the physical properties of the star.

\begin{figure}
\figurenum{5}
\hskip 0.3in \includegraphics[scale=1.0]{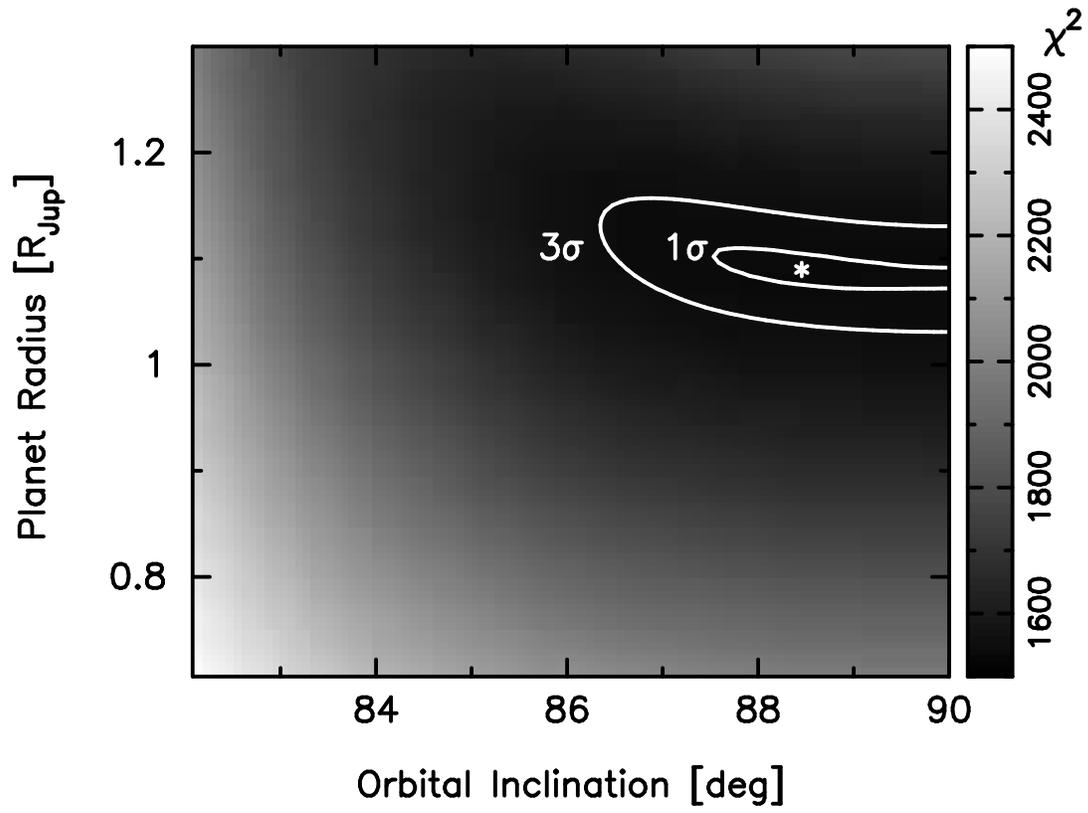}
 \caption{$\chi^2$ surface corresponding to the light curve solution
for OGLE-TR-113, in the plane of orbital inclination vs.\ planet
radius.  The number of degrees of freedom in the fit is 1512.
\label{fig:chi2}}
 \end{figure}

The photometric observations provide strong constraints on the model
fit and the $\chi^2$ surface has a well-defined minimum
(Figure~\ref{fig:chi2}). The statistical error on the radius of the
planet, $R_p$, is only $0.02$~R$_{\rm Jup}$. However, the dominant
contribution to the total error in $R_p$ is the uncertainty in the
stellar radius, $R_*$. For a fixed stellar mass of $M_* =
0.79$~M$_{\sun}$ the value of $R_*$ ($\sim$0.78~R$_{\sun}$) can be
determined to about $\pm$0.02~R$_{\sun}$ from stellar evolution
models. But the stellar mass itself is uncertain by
$\sim$0.06~M$_{\sun}$, and this propagates directly into $R_*$ because
the star must presumably conform to a model isochrone. When this
increased error for the stellar radius is accounted for the
uncertainty in the planet radius is $R_p = 1.09 \pm 0.10$~R$_{\rm
Jup}$. This is larger than the uncertainty reported by
\cite{Bouchy:04}, but we believe it is much more realistic.

\section{Discussion}

From the combination of the spectroscopic and photometric solutions we
have derived the key physical parameters of the planet. OGLE-TR-113b
has an orbital period of only 1.43 days, a mass of $1.08 \pm
0.28$~M$_{\rm Jup}$, and a radius of $1.09 \pm 0.10$~R$_{\rm Jup}$
(Table~\ref{tab:results}).  These values are consistent at the
1-$\sigma$ level with the determinations by \cite{Bouchy:04}.  Perhaps
the most interesting parameter in this case is the very short orbital
period.  OGLE-TR-56b, OGLE-TR-113b, and also the recently announced
OGLE-TR-132b \citep[][$P_{\rm orb} = 1.69$~days]{Bouchy:04} all have
orbital periods much shorter than the apparent cutoff of close-in
giant planets at around 3-day periods, determined from the radial
velocities surveys. Thus, OGLE-TR-56b (the first of these discoveries)
can no longer be considered an oddity among the extrasolar planets,
and it appears these new cases point toward an extremely interesting
new class of ``very hot'' Jupiters.

It is worth pointing out that these three short-period planets are the
result of just the first two campaigns conducted by the OGLE team, in
relatively small fields toward the Galactic center and Carina. If
these or similar surveys were to continue producing candidates at the
current rate for a period of operation similar to that of the Doppler
surveys, it is not unreasonable to expect that the number of very hot
Jupiters could increase significantly and even exceed the number of
3-4-day period planets from the radial velocity searches. The
\emph{frequency} of occurrence of very hot Jupiters, however, appears
to be much lower than that of the shortest-period Doppler planets, as
discussed by \cite{Bouchy:04}. Thus, the apparent inconsistency with
the lack of any Doppler discoveries having periods as short as those
of the OGLE planets may simply be due to a combination of their lower
rate of occurrence and the much higher sensitivity to these objects in
the photometric surveys. The latter is the result of the relatively
short duration of the OGLE photometric campaigns (a few weeks) and the
increased probability of transits from geometry, such that the chance
of finding longer-period transiting planets actually falls off
dramatically beyond $P_{\rm orb}$ of 3-4 days. The extreme conditions
of proximity to the parent stars in these very hot Jupiters opens up
the possibility of very interesting theoretical studies into their
structure and evolution, as well as migration scenarios.

\acknowledgements

MK\ acknowledges partial support by the Polish Committee for
Scientific Research, Grant No.~2P03D~001~22..  GT\ acknowledges
support for this work from the NASA's Kepler mission, STScI program
GO-9805.02-A, and the Keck PI Data Analysis Fund (JPL 1257943).  
AU was partly supported by the Polish KBN grant 2P03D02124 and the grant
`Subsydium Profesorskie'' of the Foundation for Polish Science.
GP, MTR, WG and DM gratefully acknowledge support for this research 
from the Chilean Center for Astrophysics FONDAP 15010003.
SJ\ thanks the Miller Institute for Basic Research in Science at UC
Berkeley for support through a research fellowship. We are gratefull
for a generous telescope time allocation and support
at the Las Campanas Observatory. This research has
made use of NASA's Astrophysics Data System Abstract Service.

\newpage

\newpage

%
%
 
\begin{deluxetable}{cccc}
\tablenum{1}
\tablewidth{23pc}
\tablecaption{Radial velocities measurements for OGLE-TR-113, in the
barycentric frame.\label{tab:rvs}}
\tablehead{
\colhead{HJD} & \colhead{} & \colhead{Velocity} & \colhead{Error\tablenotemark{a}} \\
\colhead{(2,400,000+)} & \colhead{Phase} & \colhead{($\kms$)} & \colhead{($\kms$)} }
\startdata
52690.6259 & 0.683 & -7.797  &  0.105 \\
52691.7230 & 0.449 & -8.008  &  0.114 \\
52692.7873 & 0.192 & -8.195  &  0.118 \\
52692.8128 & 0.210 & -8.257  &  0.116 \\
52693.7298 & 0.850 & -7.837  &  0.122 \\
52694.7927 & 0.592 & -7.707  &  0.104 \\
52754.7244 & 0.430 & -7.845  &  0.127 \\
\enddata
\tablenotetext{a}{Internal errors have been scaled to provide a
reduced $\chi^2$ of unity in the orbital solution.}
\end{deluxetable}

%
%

\begin{deluxetable}{lc}
\tablenum{2}
\tablewidth{32.4pc}
\tablecaption{Orbital and physical parameters for OGLE-TR-113b.\label{tab:results}}
\tablehead{
\colhead{\hfil ~~~~~~~~~~~~~~~~~~~~~Parameter~~~~~~~~~~~~~~~~~~~~~~} &  \colhead{Value} }
\startdata
\vspace{2pt}
~~~Orbital period (days)\dotfill                        & 1.4324758~$\pm$~0.0000046   \\
~~~Transit epoch (HJD$-$2,400,000)\dotfill & 52325.79823~$\pm$~0.00082\phm{2222}       \\
~~~Center-of-mass velocity (km~s$^{-1}$)\dotfill    & \phm{$1$}$-$7.939~$\pm$~0.043\phm{$-3$} \\
~~~Eccentricity (fixed)\dotfill                 &  0              \\
~~~Velocity semi-amplitude (m~s$^{-1}$)\dotfill        &  229~$\pm$~58\phn              \\
\vspace{10pt}
~~~Inclination angle (deg)\dotfill        &  88.4~$\pm$~2.2\phn  \\
~~~Stellar mass (M$_{\sun}$) (adopted) \dotfill  &  0.79~$\pm$~0.06 \\
~~~Stellar radius (R$_{\sun}$) (adopted) \dotfill  &  0.78~$\pm$~0.06 \\
\vspace{10pt}
~~~Limb darkening coefficient ($I$ band)\dotfill  & 0.586~$\pm$~0.015 \\
~~~Planet mass (M$_{\rm Jup}$)\dotfill          &  1.08~$\pm$~0.28      \\
~~~Planet radius (R$_{\rm Jup}$)\dotfill        &  1.09~$\pm$~0.10      \\
~~~Planet density (g~cm$^{-3}$)\dotfill     &   1.0~$\pm$~0.4 \\
~~~Semi-major axis (AU)\dotfill                  &  0.02299~$\pm$~0.00058           \\
\enddata
\end{deluxetable}

\end{document}